\documentclass[prl,twocolumn,amsmath,amssymb,superscriptaddress,longbibliography,nofootinbib]{revtex4-1}

\usepackage{graphicx}
\usepackage{bm}
\usepackage{amssymb,amsmath}
\usepackage{mathrsfs}
\usepackage{latexsym}
\usepackage{color}
\usepackage[normalem]{ulem} 
\usepackage{dcolumn}
\usepackage[colorlinks=true,citecolor=blue,urlcolor=blue]{hyperref}
\usepackage[utf8]{inputenc}
\usepackage[usenames,dvipsnames]{xcolor}

\allowdisplaybreaks

\newcommand{\CITA}{\affiliation{Canadian Institute for Theoretical Astrophysics, 60 St. George Street, Toronto, ON, M5S 3H8}}

\newcommand{\spec}{\texttt{SpEC}\ }

\begin{document}

\title{Redshift factor and the first law of binary black hole mechanics in numerical simulations}

\author{Aaron Zimmerman}
\CITA
\author{Adam G.~M.~Lewis}
\CITA
\author{Harald P.~Pfeiffer}
\CITA
\date{\today}

\begin{abstract}
The redshift factor $z$ is an invariant quantity of fundamental interest in post-Newtonian and self-force descriptions of compact binaries. 
It connects different approximation schemes, and plays a central role in the first law of binary black hole mechanics, which links local quantities to asymptotic measures of energy and angular momentum in these systems. 
Through this law, the redshift factor is conjectured to have a close relation to the surface gravity of the event horizons of black holes in circular orbits. 
We propose and implement a novel method for extracting the redshift factor on apparent horizons in numerical simulations of quasicircular binary inspirals. 
Our results confirm the conjectured relationship between $z$ and the surface gravity of the holes and that the first law holds to a remarkable degree for binary inspirals. 
The redshift factor enables tests of analytic predictions for $z$ in spacetimes where the binary is only approximately circular, giving a new connection between analytic approximations and numerical simulations.
\end{abstract}

\maketitle

{\it Introduction}.---
The relativistic two body problem is of fundamental importance in both general relativity and the astrophysics of compact objects.
Compact binaries emit gravitational radiation and inspiral, eventually merging in a dynamic, nonlinear process.
These mergers are the most promising sources of gravitational waves and provide a window into untested regimes of physics.
The landmark detection of binary black hole (BH) mergers through gravitational waves~\cite{Abbott:2016blz,Abbott:2016nmj,TheLIGOScientific:2016pea} highlights both the sophistication of waveform models and the need for further improvements to search for and interpret gravitational wave signals.
Current methods include post-Newtonian (PN) expansions in the slow velocity regime~\cite{Blanchet:2013haa}, self-force (SF) approximations~\cite{Poisson:2011nh} for systems with high mass ratios, and direct numerical solutions~\cite{Pfeiffer:2012pc,Hinder:2013oqa,Choptuik:2015mma} of inspirals beginning tens of orbits before merger.
Each method has its limitations, and they are combined into effective one body (EOB)~\cite{Buonanno:1998gg,Buonanno:2000ef} and phenomenological waveform models~\cite{Hannam:2013oca}.
In addition, connections and comparisons between the different approaches yield new insights into each of them~\cite{LeTiec:2011bk,Tiec:2014lba}.
Such insights deepen our understanding of relativity and maximize the scientific benefits of future gravitational wave observations.

Invariant quantities play a crucial role in these comparisons, since each method 
uses different gauges and various approximation schemes.
The invariant redshift factor $z$ has proven essential in comparisons between analytic approximations, as first discussed for 
circular binaries~\cite{Detweiler:2008ft}.
Such systems remain stationary in the corotating frame, having a helical symmetry embodied in a helical Killing vector (HKV) $K^\mu$.
In this context, the redshift factor allows for comparison of results obtained in distinct coordinate gauges~\cite{Sago:2008id,Shah:2010bi}, and has played a central role in the development of PN and EOB theory using SF, e.g.~Refs.~\cite{Blanchet:2010zd,LeTiec:2011ab,Bini:2013zaa}.

For isolated BHs, the laws of black hole mechanics are relations between the area, angular momentum, and charge of the hole~\cite{Bekenstein:1973ur,Bardeen:1973gs}.
These relations provide deep insights into BH spacetimes and, combined with quantum field theory, reveal the thermodynamic nature of BHs~\cite{Hawking:1974sw}.
Modified laws of BH mechanics exist in spacetimes with a HKV, interrelating properties of the orbiting bodies (stars or BHs)~\cite{Friedman:2001pf}.
There is a compelling connection between the redshift factor $z$ and these laws: when applied to HKV spacetimes in the PN and SF approximations, where compact objects are represented as test bodies, $z$ enters the laws as a generalized force, in direct analogy to the surface gravity of the object~\cite{LeTiec:2011ab}.
With the help of $z$, these laws play a growing role in PN~\cite{LeTiec:2011ab,Blanchet:2012at}, SF~\cite{Gralla:2012dm,Isoyama:2014mja} and EOB~\cite{Bini:2013rfa,Bini:2013zaa,Bini:2016cje} modeling of binaries.
These models inform theoretical templates used by Advanced LIGO~\cite{TheLIGOScientific:2014jea} to enable detection and characterization of gravitational waves emitted by compact binaries~\cite{Abbott:2016blz,TheLIGOScientific:2016qqj,TheLIGOScientific:2016wfe}.

The extension of redshift-based analyses to numerical simulations faces two problems.
First, past comparisons have taken place in the context of conservative dynamics, where a HKV exists exactly or in an averaged sense for eccentric orbits.
However, in simulations dissipation is present, causing the BHs to inspiral and breaking helical symmetry.
Second, in analytic theory $z$ is computed on particle worldlines, but simulations deal with extended bodies whose interiors may be excised from the computational domain, so that no worldline is available.
Thus, while $z$ has been used extensively to communicate between analytic methods, it has not been used in simulations.
An exception is the study of a related connection between the Bondi energy $E_{B}$ and angular momentum $J_{B}$ in simulations~\cite{Damour:2011fu,LeTiec:2011dp,Nagar:2015xqa}.

This Letter reports the first application of redshift comparisons to numerical simulations.
While quasicircular binaries do not possess a strict HKV, their early inspiral can be approximated by adiabatic evolution through a sequence of circular orbits. 
As such, they possess an approximate HKV.
We discuss how the connection between the $z$ and the surface gravity on BHs in the presence of a HKV yields a normalized surface gravity and corresponding $z$ for BHs.
We then extract $z$ from nonspinning binary BH simulations and compare it to PN results.
Finally, with our redshift factor we test the laws of binary BH mechanics for quasicircular orbits.
We also validate the conjecture that the first law holds when dissipation is present, when formulated in terms of Bondi quantities~\cite{LeTiec:2011ab}.

{\it The first law of binary black hole mechanics and the redshift factor}.---
Consider a spacetime with HKV $K^\mu$ containing two BHs.
When the spacetime is asymptotically flat, we write $K^\mu=(\partial_t)^\mu+\Omega(\partial_\phi)^\mu$, where $t$ is the asymptotic Killing time measured by inertial observers, $(\partial_\phi)^\mu$ is a spacelike vector with closed orbits of length $2 \pi$, 
and $\Omega$ is the orbital frequency of the Killing flow.
The BHs have Killing horizons, with the tangents to their generators $\chi^\mu$ equal to $K^\mu$.
The surface gravity $\kappa$ of the hole is defined by $\chi^\mu\nabla_\mu\chi^\nu=\kappa\chi^\nu$, and is constant on the horizon.
We see that $\kappa$ arises from the nonaffine parameterization of $\chi^\mu$, which is normalized to equal $K^\mu$. 
The normalization of $K^\mu$ is fixed by the vector $(\partial_t)^\mu$, hence by an asymptotic inertial frame.
In the case of a single BH, $(\partial_t)^\mu$ and $(\partial_\phi)^\mu$ are the Killing vectors of Kerr and $\Omega$ is the horizon frequency.

HKV spacetimes obey the first law of binary mechanics, which governs the variations of a Noether charge $Q$~\cite{Wald:1993nt,Friedman:2001pf},
\begin{align}
\label{eq:DiffFL}
\delta Q & =\delta M  - \Omega \delta J  = \kappa_1 \frac{\delta A_1}{8\pi} + \kappa_2 \frac{\delta A_2}{8\pi} \,.
\end{align}
Here $M$ and $J$ are the Arnowitt-Deser-Misner (ADM) mass and angular momentum of the spacetime, and $\kappa_i$ are the surface gravities of the BHs.
The power of \eqref{eq:DiffFL} lies in the connection of quantities defined on the BHs to asymptotic quantities, through the global vector field $K^\mu$.
Strictly speaking, both \eqref{eq:DiffFL} and asymptotic flatness require that some conservative approximation to general relativity holds in the HKV spacetime.

Next, consider spacetimes with point particles on circular orbits. 
These particles model compact objects in PN and SF approximations. 
The redshift factor $z$ of a particle moving with four-velocity $u^\mu$ is defined as $z=1/u^t$,
where $u^t=(\partial_t)^\mu{}u_\mu$. 
In a certain effective metric, $z$ compares the clock rates at the particle and at infinity, or equivalently the redshift of light emitted perpendicular to the particle's motion.
In this sense $z$ is a well-defined observable for the particle, and is preserved by helically symmetric gauge transforms~\cite{Detweiler:2008ft}.

Reference~\cite{LeTiec:2011ab} derives a modified law of BH mechanics for nonspinning point particles, $\delta Q=z_1\delta m_1+z_2\delta m_2$, with $z_i$ and $m_i$ the redshift factors and masses of the particles.
Here, $z_i$ takes the role of the surface gravity of the BH it replaces.
These relations have been verified to high PN order and leading order in SF for corotating systems~\cite{Gralla:2012dm}, and have been used to develop analytic approximations and waveform models, e.g.~\cite{Bini:2013rfa,Bini:2013zaa,Bini:2016cje,Bernard:2015njp,Damour:2016abl}, often by assuming they continue to hold at higher orders.
The variational equations imply an integral relation~\cite{Friedman:2001pf,LeTiec:2011ab}
\begin{align}
\label{eq:FL}
Q = M - 2 \Omega J = z_1 m_1 + z_2 m_2 = \kappa_1 \frac{A_1}{4\pi} + \kappa_2 \frac{A_2}{4\pi} \,,
\end{align}
for spacetimes containing point particles or BHs, respectively.
This relation connects local notions of the surface gravity to the redshift factor and the energy and angular momentum of the spacetime.

\begin{figure}[t]
\includegraphics[width=0.98\columnwidth]{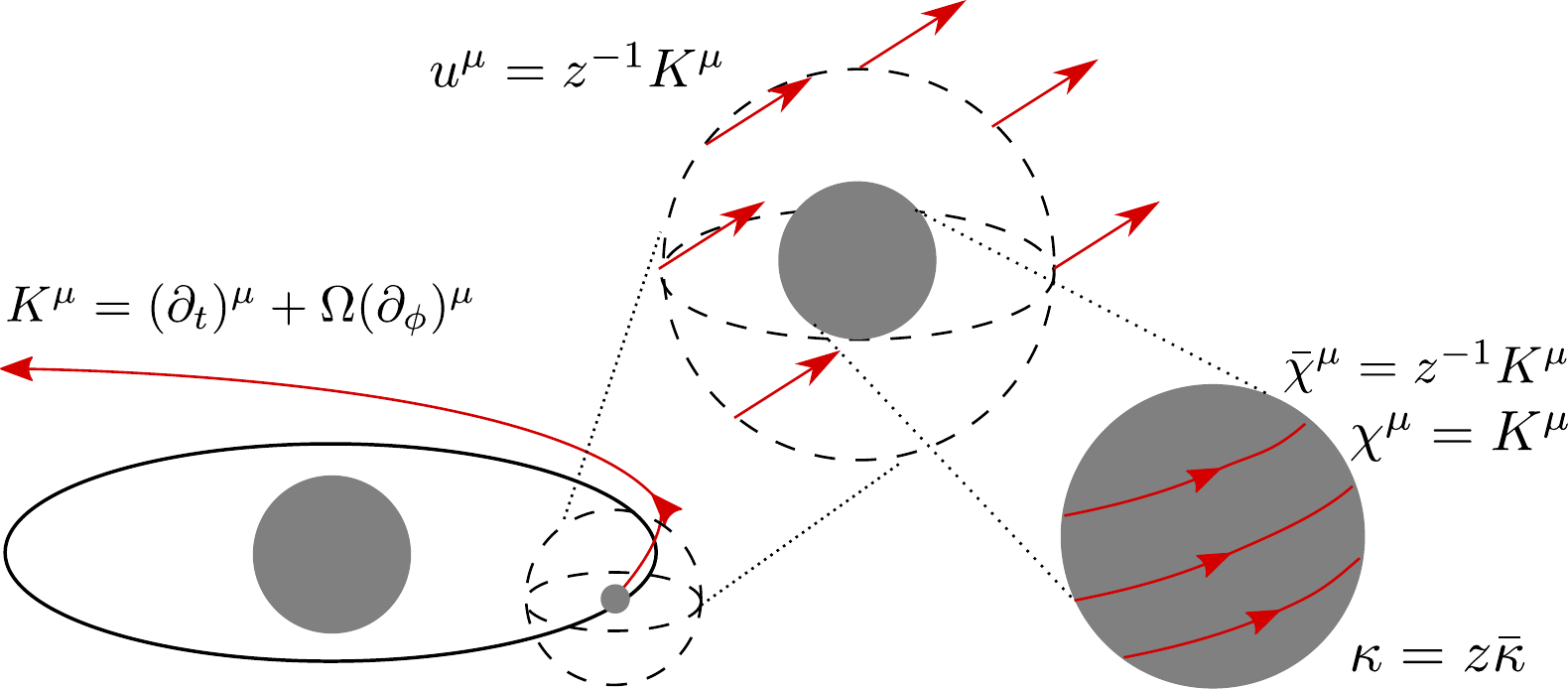} 
\caption{Illustration of the connection between the redshift factor $z$ and the surface gravity $\kappa$ of a small black hole, using a matched asymptotic analysis of a HKV spacetime.}
\label{fig:RedshiftCartoon}
\end{figure}

{\it Surface gravity and the redshift factor}.---
Our first step is to make sense of $z$ in a spacetime without a particle worldline on which to evaluate $u^t$. 
The first law, Eq.~\eqref{eq:FL}, connects the surface gravity and redshift factor: if we equate the masses $m_i$ to the irreducible masses of the holes, $m_{i}=\sqrt{A_i/(16\pi)}$, then $\kappa_i=z_i/(4m_{i})$ and we have
\begin{align}
\label{eq:SGRelation}
\kappa_i  = z_i \bar \kappa_i \,,
\end{align}
where $\bar\kappa$ is the surface gravity of an isolated BH.
In the limit of infinite separation, $z_i \to 1$, and $\kappa_i$ reduce to the expected values. 

A heuristic derivation of Eq.~\eqref{eq:SGRelation} is given in Fig.~\ref{fig:RedshiftCartoon}, which depicts a matched asymptotic picture of a BH binary with masses $m_1$, $m_2$, and small mass ratio $q=m_2/m_1$. 
The smaller hole is surrounded by a matching region which remains large compared to it, but becomes arbitrarily small in the limit $q\ll1$.
In the matching zone, we consider a family of comoving observers with four-velocity $u^\mu$; this velocity field must be parallel to $K^\mu$ by the symmetry, and becomes equal to the point particle velocity in the limit $q\ll1$.
Because $u^t=z^{-1}$ we have $u^\mu=z^{-1}K^\mu$.
From the perspective of these observers, the small BH is an isolated BH immersed in an external tidal field~\cite{Detweiler:2000gt}.
The observers use their own asymptotic normalization of the HKV to define the tangents $\bar\chi^\mu=z^{-1}\chi^\mu$, and define their rescaled surface gravity through $\bar\chi^\mu\nabla_\mu\bar\chi^\nu=\bar\kappa\bar\chi^\nu$.
The key idea is that $\bar\kappa$ is in fact the surface gravity of an isolated BH; tidal corrections scale with the square of $m_2$ and are negligible in the test particle limit.
All of the above considerations hold for comparable mass systems so long as the radius of curvature $\mathcal{R}$ due to external influences on each of the holes is large compared to the size of the hole, $\mathcal{R}\gg m_i$.

Equation~\eqref{eq:SGRelation} can be made rigorous in a HKV spacetime using matched asymptotics~\cite{PoundRedshift:2015} and is straightforward to demonstrate for isolated boosted BHs and BHs immersed in an axisymmetric external potential.
Equation~\eqref{eq:SGRelation} allows us to compare analytic predictions of $z$ to $\kappa$ of the corresponding BH in a simulation, although we expect it will begin to break down when the system becomes very relativistic, except on the smaller hole when $q\ll1$.

Equation~\eqref{eq:SGRelation} relies on the normalization of $\chi^\mu$ in terms of $K^\mu$ and the asymptotic observers, in particular $\chi^t = 1$.
This normalization is not available in a numerical spacetime.
We only measure tangents $l^\mu$ with some unknown normalization, with $\kappa_{(l)}$ given by $l^\mu\nabla_\mu l^\nu=\kappa_{(l)}l^\nu$.
We have $l^\mu=\alpha\chi^\mu$ for some factor $\alpha$.
The surface gravity inherits this rescaling, $\kappa_{(l)}=\alpha \kappa$.
The unknown $\alpha$ therefore cancels out of the ratio $\kappa_{(l)}/l^t=\kappa/\chi^t=\kappa$.
Using Eq.~\eqref{eq:SGRelation} we arrive at an expression for $z$ which is invariant under a rescaling of $l^\mu$,
\begin{align}
\label{eq:NumRedshift}
z & = \frac{\kappa_{(l)}}{l^t \bar \kappa}  \,.
\end{align}
This also accounts for time transformations of the form $t \to \tilde t (t)$. 

Simulations usually track apparent horizons (AHs) rather than the event horizon (EH).
The AHs are good approximations to the EH until near merger~\cite{Cohen:2008wa}. 
As such, we use the outward null normals $l^\mu$ to the AHs to evaluate Eq.~\eqref{eq:NumRedshift}.
We compute $z$ pointwise on each AH and horizon-average $z$ for our final result.
This allows for a different rescaling of each $l^\mu$, and mitigates any tidal effects contaminating $\kappa$.

{\it Approximate helical symmetry and expected errors}.---In order to make any sense of Eq.~\eqref{eq:FL} in quasicircular inspirals, we must assume that this relation holds for approximate HKVs.
We imagine that the binary inspirals adiabatically, passing from circular orbit to circular orbit.
At each stage, a small region of spacetime enclosing the binary can be approximated by a HKV spacetime.
The boundary of each region can be connected to asymptotic infinity by a null surface, and radiation propagating on this surface inherits the approximate HKV.
It is clear from these considerations that we should use the Bondi mass and angular momentum in Eq.~\eqref{eq:FL}, which are constant on each asymptotic null surface but vary as the inspiral proceeds~\cite{LeTiec:2011ab}. 
Making this argument rigorous using a two-time-scale expansion~\cite{Hinderer:2008dm,Pound:2015wva} is an open problem, and our results provide evidence that this can be done.

With this in mind, we can estimate the sources of error which prevent Eq.~\eqref{eq:FL} from holding exactly.
Killing's equation is violated for the approximate HKV: $\nabla_{(\mu}l_{\nu)}\neq0$.
This generates a shear $\sigma_{\mu\nu}$ of the null generators, which represents gravitational waves entering the horizon and increasing its mass~\cite{Hawking:1972hy}, so that $\dot{m}\sim|\sigma|^2$.
In our simulations $\dot{m}\sim10^{-9}$ and so $|\sigma|\sim10^{-5}$.
Conservatively, nonadiabatic effects can be expected to scale as $\dot{\Omega}/(2 \Omega^2)$ which is typically $\sim10^{-2}-10^{-3}$ until near merger.
Furthermore, only corotating binaries have a strict HKV, but corotation cannot occur for an inspiral where $\Omega$ evolves. 
We naively require $\Omega_H \sim \Omega$, and errors for our nonspinning configurations scaling as $\Omega_H^2\sim\Omega^2 \sim10^{-2}-10^{-4}$, although studies of initial configurations indicate a smaller error in practice~\cite{Caudill:2006hw}. 
Using AHs rather than the EH introduces another source of error during the final plunge and merger.
Before the plunge, we expect the AH and EH to be identical, since the EH generators approach the AH exponentially moving backward in time from merger, with $e$-folding time $1/\kappa_i$~\cite{Cohen:2008wa}.
The AHs are not very dynamic before merger, and so the generators have no difficulty reaching the AH.
We leave a full investigation and possible mitigation of these errors for future study.

{\it Numerical simulations}.---We simulate quasicircular inspirals using the \spec code~\cite{SpECwebsite} for three mass ratios $q =1$, $2/5,$ and $2/7$, in order to explore the dependence of $z$ on $q$.
We use five resolutions for the $q=1$ case, and two for each of the $q=2/5$ and $q=2/7$ cases.
Our parameters are chosen to give circularized~\cite{Mroue:2010re,Buonanno:2010yk}  binaries previously presented in Ref.~\cite{Chu:2015kft}, or from the publicly available catalog~\cite{SXSCatalog} reported in Ref.~\cite{Mroue:2013xna}.
These BH binaries execute $\sim28$ orbits, beginning at initial orbital frequencies of $m\Omega_0 \times 10^{2}\approx 1.22,1.33,$ and $1.46$ for $q=1$, $2/5$ and $2/7$, where $m$ is the total mass of the holes.
They have initial eccentricities $e < 10^{-4}$.
Our numerical error, represented by the difference in numerical values across resolutions, is always much smaller than the difference between the numerical and analytic results.

\begin{figure}[t]
\includegraphics[width=0.98\columnwidth]{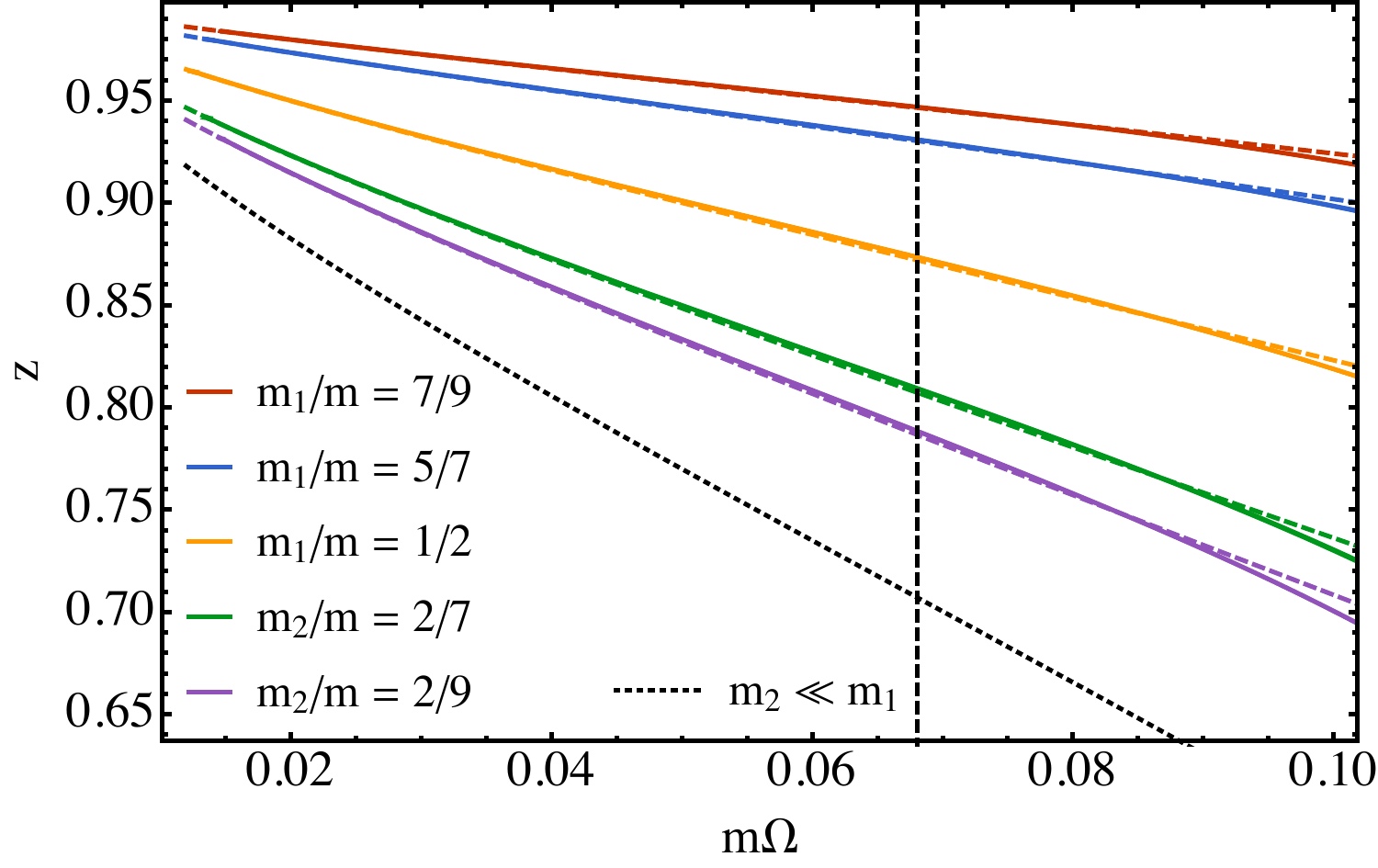} 
\caption{Redshift plotted for three inspirals with $q=1,2/5,$ and $2/7$ (solid), with all resolutions shown (although they are not distinguishable), together with 3PN (dashed) and test particle predictions (dotted). We label the curves by the masses of each BH, in units of total mass $m=m_1+m_2$.}
\label{fig:zPlots}
\end{figure}

We measure the orbital frequency using extrapolated waveforms~\cite{Boyle:2008ge,Boyle:2009vi,Boyle:2013nka}, by defining $\Omega(t) = \omega_{22}(t_r)/2$, where $\omega_{22}$ is the frequency of the $l=m=2$ mode of the waveform. 
References~\cite{Taylor:2013zia,Chu:2015kft} compared waveform extrapolation to the more sophisticated procedure of Cauchy characteristic extraction~\cite{Bishop:1996gt,Babiuc:2010ze}, indicating that systematic errors introduced by extrapolation are small enough to neglect here.
Local and asymptotic quantities are compared at equal retarded times $t_r = t - r_*$, with $r_*$ a tortoise coordinate defined with respect to the ADM mass of the spacetime~\cite{Boyle:2009vi}.
Using $t_r$ to propagate the asymptotic frequency to the location of the BHs captures most of the expected relativistic effects, but given the precision of our comparison further study of this matching is warranted.
Local and asymptotic quantities $f(\Omega)$ are compared to PN predictions at equal orbital frequencies, $\Omega_{\rm orb} = \Omega$; we neglect the relative 5PN differences between $\Omega_{\rm orb}$ and $\omega_{22}/2$ for PN binaries.
The extrapolated waveform is used to compute the energy and angular momentum fluxes from the simulation, and the Bondi quantities are computed by subtracting the integrated flux~\cite{Boyle:2008ge} from the initial ADM mass and angular momentum of the spacetime, $M_{B}(t_r)=M_{\rm ADM}-\int_0^{t_r}dt\,\dot{E}$, and similarly for $J_{B}$. 

We measure the redshift factor in Eq.~\eqref{eq:NumRedshift} using the outward null normals $l^\mu$ of the AHs, and compute $m_{i}$ using the area of the AHs. 
We monitor the spins of the BHs computed using the approximate Killing vector method~\cite{Lovelace:2014twa}, and they remain negligible, $S_i/m_i^2\lesssim 10^{-5}-10^{-4}$, until very near merger.

\begin{figure}
\includegraphics[width=0.98\columnwidth]{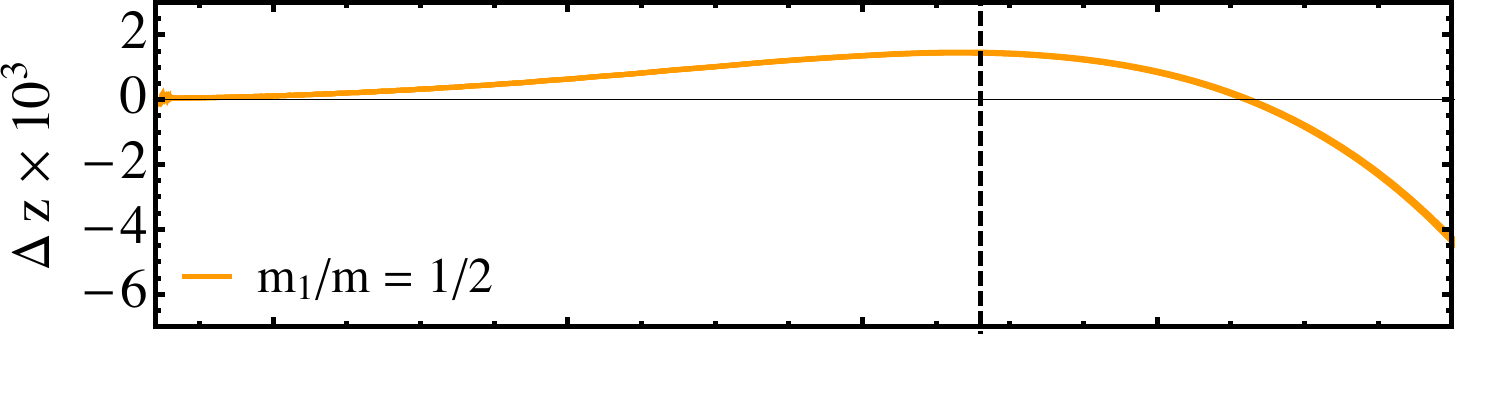} \\
\vspace{-12pt}
\includegraphics[width=0.98\columnwidth]{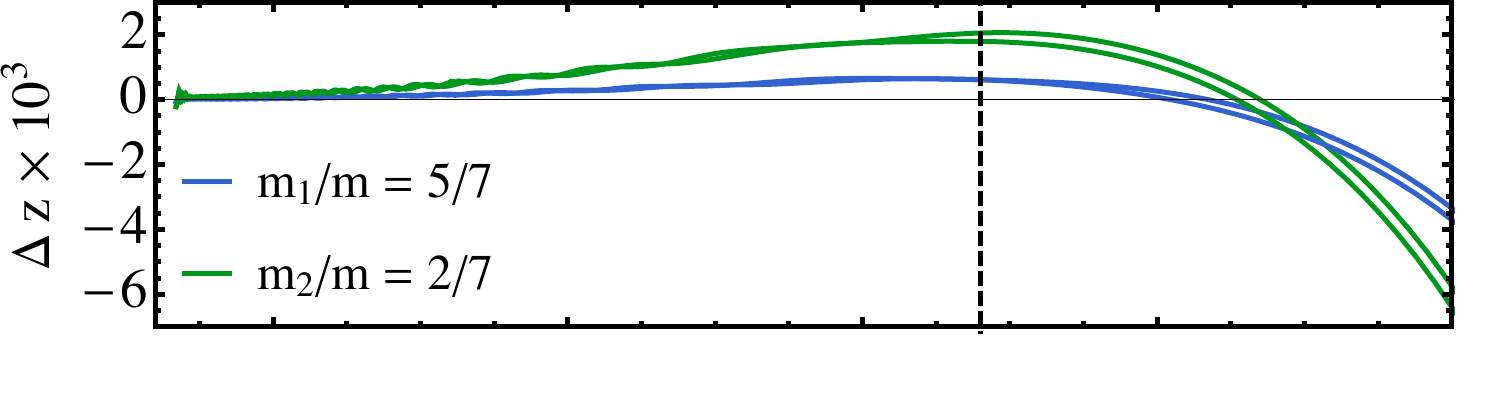} \\
\vspace{-12pt}
\includegraphics[width=0.98\columnwidth]{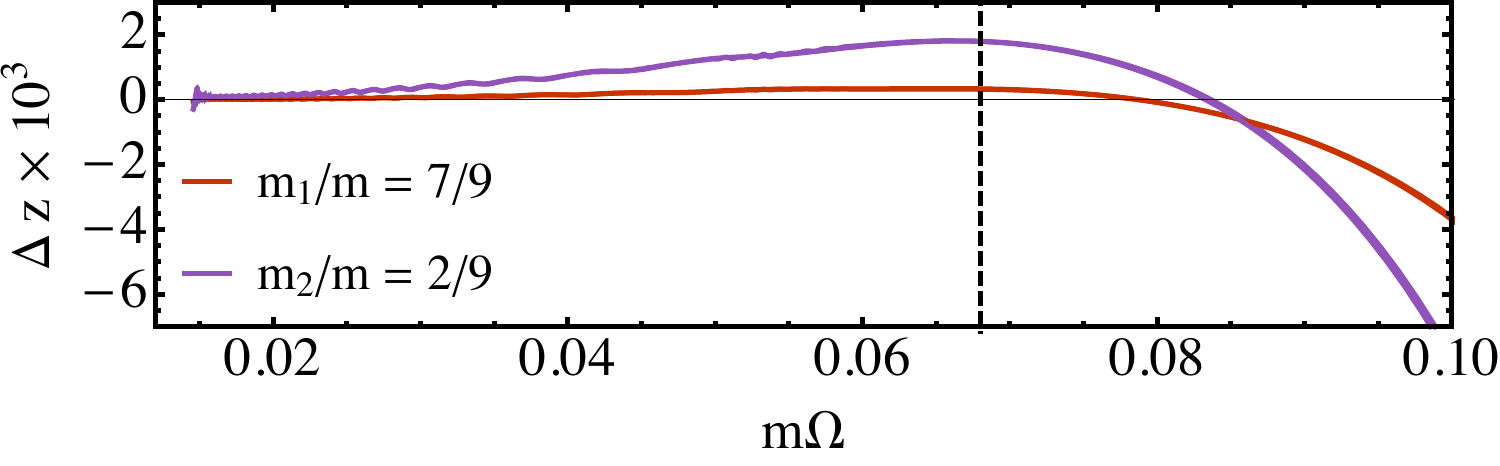} 
\caption{Differences $\Delta z$ between PN predictions and numerical values of $z$ in Fig.~\ref{fig:zPlots}, multiplied by $10^3$, for our three binaries: $q=1$ (top), each member of  $q=2/5$ (middle) and each member of $q =2/7$ (bottom). All resolutions are plotted.}
\label{fig:zDiffPlots}
\end{figure}

{\it Results}.---Figure~\ref{fig:zPlots} plots $z(m\Omega)$ for all resolutions of our three binary systems together with the 3PN analytic prediction of Ref.~\cite{LeTiec:2011ab} and the test particle limit. 
We indicate the innermost stable circular orbit (ISCO) frequency for a Schwarzschild BH of total mass $m$ by the vertical dashed line in all our plots, which we cut off shortly before a common AH forms.
The general trends can be understood by considering the test particle limit $z = \sqrt{1 - 3 (m \Omega)^{2/3}}$, and we see that the less massive BHs have a stronger redshift effect due to the deeper gravitational potential of the more massive holes.

Since $z$ is approximately unity, in Fig.~\ref{fig:zDiffPlots} we simply show the difference $\Delta z=z_{\rm{NR}}-z_{\rm{PN}}$ between numerical and PN predictions for $z$.
We see remarkable agreement with the PN predictions in all cases even through the beginning of the plunge, with better agreement for more massive BHs. 
The small oscillations seen for $q=2/5$ and $2/7$ arise from residual eccentricities, which are a factor $\sim4$ larger than for $q=1$.
Our numerical errors are always several times smaller than $\Delta z$.
Note that the difference between the 2PN and 3PN predictions are also quite small for these binaries, $\lesssim\Delta z$, so we are likely probing the nonadiabatic effects discussed above.

\begin{figure}
\includegraphics[width=0.98\columnwidth]{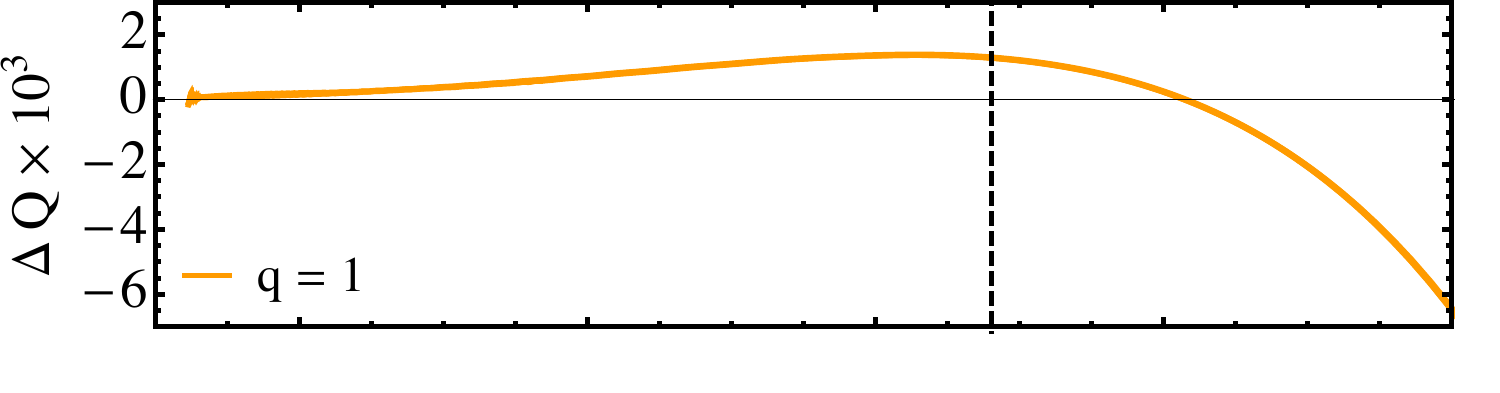} \\
\vspace{-12pt}
\includegraphics[width=0.98\columnwidth]{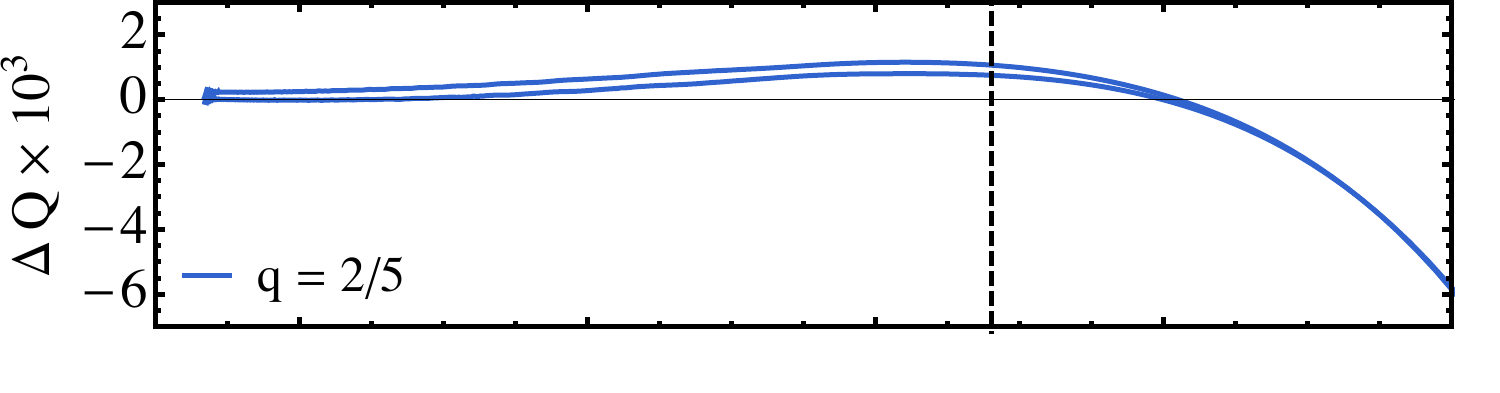}\\
\vspace{-12pt}
\includegraphics[width=0.98\columnwidth]{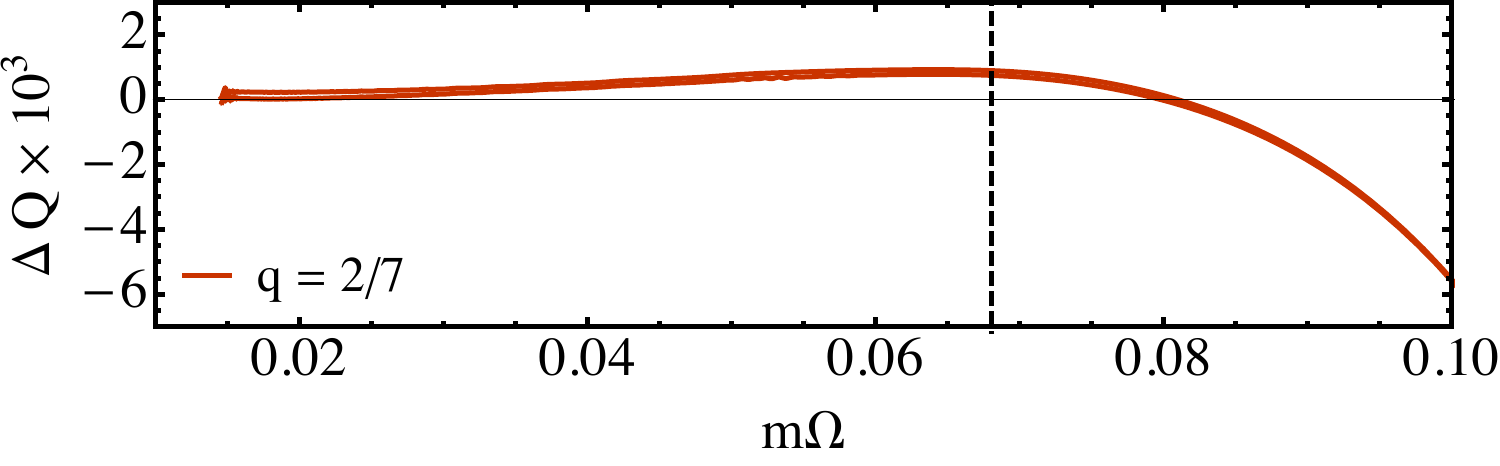} 
\caption{Deviation $\Delta Q$ from the first law, Eq.~\eqref{eq:FL}, multiplied by $10^3$, for our three binaries. 
All resolutions are plotted.}
\label{fig:FLDiffPlots}
\end{figure}

Finally, we test the first law, Eq.~\eqref{eq:FL}. Figure~\ref{fig:FLDiffPlots} plots the difference $\Delta Q=z_{1}m_{1}+z_{2}m_{2}-(M_{\rm{B}}-2\Omega J_{\rm{B}})$,
which is dominated by $\Delta z$ in all cases, resulting in similar deviations as in Fig.~\ref{fig:zDiffPlots}.
We emphasize that this comparison is sensitive to many aspects of the conjectured relation~\eqref{eq:FL} as applied to binary inspirals, including the mapping at retarded times between local and global quantities, contamination by nonadiabatic effects, and the use of Bondi quantities in the first law.
In this sense, the close agreement is remarkable, and may improve further if those errors that can be controlled are dealt with.
This agreement even during the plunge indicates that a modified first law may apply beyond the slow inspiral regime studied by analytic approximations.

The redshift factor of the merged BH approaches unity when $\bar \kappa$ is taken to be that of a Kerr BH with the final mass and spin.

{\it Discussion}.---We have presented the first extraction of the redshift factor $z$ from simulations of BH inspirals, by exploiting a connection between $z$ and the appropriately normalized surface gravity of the hole.
The result is in good agreement with PN theory for several mass ratios. 
We have tested the first law of binary black holes in the nonadiabatic regime for the first time, finding remarkable agreement.

Our results are the first step towards a variety of future connections between simulations and analytic methods. 
We will next investigate higher mass ratios to test SF predictions and extract higher order SF terms.
Following this, we can explore spinning~\cite{Blanchet:2012at,Shah:2012gu} and eccentric~\cite{Tiec:2015cxa,vandeMeent:2015lxa} binaries, where modified first laws hold.
A second direction of study is to investigate the redshift factor on the actual EHs of BH spacetimes, although this requires intensive postprocessing~\cite{Cohen:2011cf,Bohn:2016afc}.
One could also improve the extraction of the redshift by developing a method to compute the best approximate HKV at each time step, analogous to the method used in \spec to compute BH spins~\cite{Cook:2007wr,Lovelace:2008tw,Beetle:2008yt,Lovelace:2014twa}.
Finally, our results motivate formal studies of spacetimes with an approximate HKV.

Looking toward the future, we envision $z$ as one of a family of invariant quantities used to interconnect analytic theory, waveform models, and numerical simulations.
As we continue to refine our understanding of the relativistic two-body problem, these insights will transfer to the understanding of gravitational wave emission from these systems, and in turn improve our ability to draw astrophysical insights from compact binaries in the nascent era of gravitational wave astronomy.

\begin{acknowledgements}
We thank Takahiro Tanaka for first suggesting this method for computing the redshift factor in numerical spacetimes. 
This work was conceived at the 18th Capra meeting and molecule workshop YITP-T-15-3 at the Yukawa Institute for Theoretical Physics. We thank the participants of this conference for valuable discussions, especially Takahiro Tanaka, Alexandre Le Tiec, Adam Pound, and Leor Barack.
We also thank  Abraham Harte, Ian Hinder, Soichiro Isoyama, and Eric Poisson for valuable discussions, Serguei Ossokine for assistance in computing the energy and angular momentum fluxes to infinity, Tony Chu for providing circularized initial data parameters for some of our simulations, and Adam Pound for valuable comments on this Letter.
Calculations were performed using the Spectral Einstein code ({\tt SpEC})~\cite{SpECwebsite}.
We gratefully acknowledge funding from NSERC of Canada, the Ontario Early Researcher Awards Program, the Canada
Research Chairs Program, and the Canadian Institute for Advanced Research;
A.~Z.~was supported by the Beatrice and Vincent Tremaine Postdoctoral Fellowship at the Canadian Institute for Theoretical Astrophysics during a portion of this work.
Computations were performed on the GPC supercomputer at the SciNet HPC Consortium~\cite{SciNet}. 
SciNet is funded by the Canada Foundation for Innovation under the auspices of Compute Canada, the Government of Ontario, Ontario Research Fund - Research Excellence, and the University of Toronto.
\end{acknowledgements}

\bibliography{RedshiftFromNRLetter}

\end{document}